\documentclass[12pt]{amsart}

\usepackage{amssymb,amsmath,amscd,amsthm,enumerate,listings,verbatim,xcolor,url,float}
\usepackage[utf8]{inputenc}
\usepackage{mathrsfs}
\usepackage{bbm}
\usepackage{stmaryrd}
\usepackage[left=1.3in,right=1.3in,top=1.2in,bottom=1.2in]{geometry}
\usepackage[colorlinks=true,linkcolor=black,anchorcolor=black,citecolor=black,filecolor=black,menucolor=black,runcolor=black,urlcolor=black]{hyperref}
\usepackage{enumitem}
\usepackage{bm}
\usepackage{tikz}\usetikzlibrary{positioning,patterns,shapes,arrows,calc}
\usepackage{arydshln}	

\DeclareFontFamily{U}{mathx}{\hyphenchar\font45}
\DeclareFontShape{U}{mathx}{m}{n}{
      <5> <6> <7> <8> <9> <10>
      <10.95> <12> <14.4> <17.28> <20.74> <24.88>
      mathx10
      }{}
\DeclareSymbolFont{mathx}{U}{mathx}{m}{n}
\DeclareFontSubstitution{U}{mathx}{m}{n}
\DeclareMathAccent{\widecheck}{0}{mathx}{"71}
\DeclareMathAccent{\wideparen}{0}{mathx}{"75}

\definecolor{purple}{rgb}{.5,0,1}
\definecolor{orange}{rgb}{1,.5,0}
\definecolor{pink}{rgb}{1,0,.5}
\definecolor{green}{rgb}{0,.5,0}
\definecolor{gold}{rgb}{1,.623,0}

\newtheorem{theorem}{Theorem}[section]

\newtheorem{prop}[theorem]{Proposition}

\theoremstyle{definition}
\newtheorem{remark}[theorem]{Remark}
\newtheorem{definition}[theorem]{Definition}

\usepackage{todonotes}

\definecolor{codegreen}{rgb}{0,0.6,0}
\definecolor{codegray}{rgb}{0.5,0.5,0.5}
\definecolor{codepurple}{rgb}{0.58,0,0.82}
\definecolor{backcolour}{rgb}{0.95,0.95,0.92}

\lstdefinestyle{mystyle}{
    backgroundcolor=\color{backcolour},   
    commentstyle=\color{codegreen},
    keywordstyle=\color{magenta},
    numberstyle=\tiny\color{codegray},
    stringstyle=\color{codepurple},
    basicstyle=\ttfamily\footnotesize,
    breakatwhitespace=false,         
    breaklines=true,                 
    captionpos=b,                    
    keepspaces=true,                 
    numbers=left,                    
    numbersep=5pt,                  
    showspaces=false,                
    showstringspaces=false,
    showtabs=false,                  
    tabsize=2
}

\begin{document}
\title{Abstract art generated by Thue-Morse correlation functions}
\author{Darren C. Ong}
\address{Department of Mathematics and Applied Mathematics, Xiamen University Malaysia, Jalan Sunsuria, Bandar Sunsuria, 43900 Sepang, Malaysia}
\email{darrenong@xmu.edu.my}

\maketitle
\begin{section}{Introduction}
Aperiodic order refers to a mathematical structures that are not periodic, but are nevertheless highly ordered and close to periodic in some way. Aperiodically ordered patterns gained increased interest among physicists and mathematicians upon the discovery of the quasicrystal in 1982, since these structures were useful in understanding the properties of quasicrystals. For an overview of aperiodic order and in particular its connection to crystallography, please consult \cite{Moody1997} \cite{baakegrimm}and \cite{baakegrimm2}.

These aperiodic ordered structures have long been associated with remarkable images. Among them are the iconic diffraction pattern of the quasicrystal with 10-fold rotational symmetry (Figure \ref{fig:diffraction}), the Penrose tiling, and the Hoftstadter butterfly, which is a graphical solution of the Harper's equation. (The Harper's equation can be interpreted as a Schr\"odinger equation with aperiodically ordered potential). 

\begin{center}
\begin{figure}[H]
\caption{Electron diffraction pattern. Image by Frederic Mompiou}
\includegraphics[scale=1]{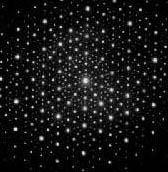}
\label{fig:diffraction}
\end{figure}
\end{center}

Interest in these aperiodically ordered patterns emerged in the arts long before they were studied by scientists. Aperiodically ordered structures appear in medieval Islamic architecture \cite{LuSteinhardt2007},\cite{alajlouni2012}, \cite{ATR2014}. German renaissance artist Albrecht D\"urer also experimented with aperiodic tilings \cite{luck2000}. Aperiodic order has also emerged in music composition \cite{Prusinkiewicz1986}, \cite{Ong2020},\cite{Trevino2022}.

In this paper, we present a novel approach to creating art from aperiodically ordered patterns. Unlike many of the examples cited above, this art is not based on aperiodically ordered tilings. Rather, it emerges from the symmetries inherent in these patterns that form the diffraction patterns similar to the ones in Figure \ref{fig:diffraction}.

This art is constructed from an infinite binary sequence known as the Thue-Morse (or Prouhet-Thue-Morse) sequence. This sequence will be defined in the next section, but for now we will just say that the first few terms are
$$
abbabaabbaababba\ldots
$$
This sequence is not periodic, but is nevertheless close to periodic in the following way. Given an infinite sequence $S$, we define its complexity sequence $\{C_n^S\}$ as follows. The entry $C_n^S$ for $n\in\mathbb Z_+$ is the number of distinct subsequences of $S$ of length $n$. If $S$ is a periodic or eventually periodic sequence, it is not hard to see that $\{C_n^S\}$ is bounded. If $S$ is a random binary sequence, almost surely $\{C_n^S\}=2^n$. According to the Morse-Hedlund theorem (e.g. Proposition 4.1 of \cite{baakegrimm}), if $S$ is not periodic then its corresponding $\{C_n^S\}$ grows at least linearly. It can be shown that if $S$ is the Thue-Morse sequence, then its corresponding $\{C_n^S\}$ grows linearly (more precisely, by Proposition 4.5 of \cite{Brlek89} $C_n^S$ is bounded above by $10n/3$). In this sense, we can say the Thue-Morse sequence is aperiodic, but as close to periodic as possible. We thus describe it as an ``aperiodically ordered" sequence.  

We will look at the \emph{autocorrelation function} corresponding to this Thue-Morse sequence. The autocorrelation function $\eta(n)$ in essence measures how similar a sequence is to a copy of itself shifted $n$ steps. It is used for calculating the diffraction pattern of a quasicrystal structure to obtain images like in Figure \ref{fig:diffraction}  For a random, independent identically distributed sequence the autocorrelation function is almost always constant. For a periodic sequence, the autocorrelation function $\eta(n)$ is also periodic. But for an aperiodically ordered sequence, $\eta(n)$ has a very complicated and interesting structure. 

For our art project, we look at a generalization, the fourth order autocorrelation function $\eta(m,n,k)$. This measures how much the Thue-Morse sequence and three copies of itself shifted $m$, $n$ and $k$ steps respectively are similar to each other. We fix the $k$, and create a matrix whose $(i,j)$ entry contains $\eta(i,j,k)$. We then assign colours to each matrix entry, creating an image in the fashion of a heatmap. As a preview, we will show an example image in Figure \ref{fig:401}. More elaborate examples will be shown later in the paper. For more on the mathematical properties of this higher order autocorrelation function, please consult \cite{BaakeCoons}.

\begin{center}
\begin{figure}[H]
\caption{Example of autocorrelation art}
\includegraphics[scale=1]{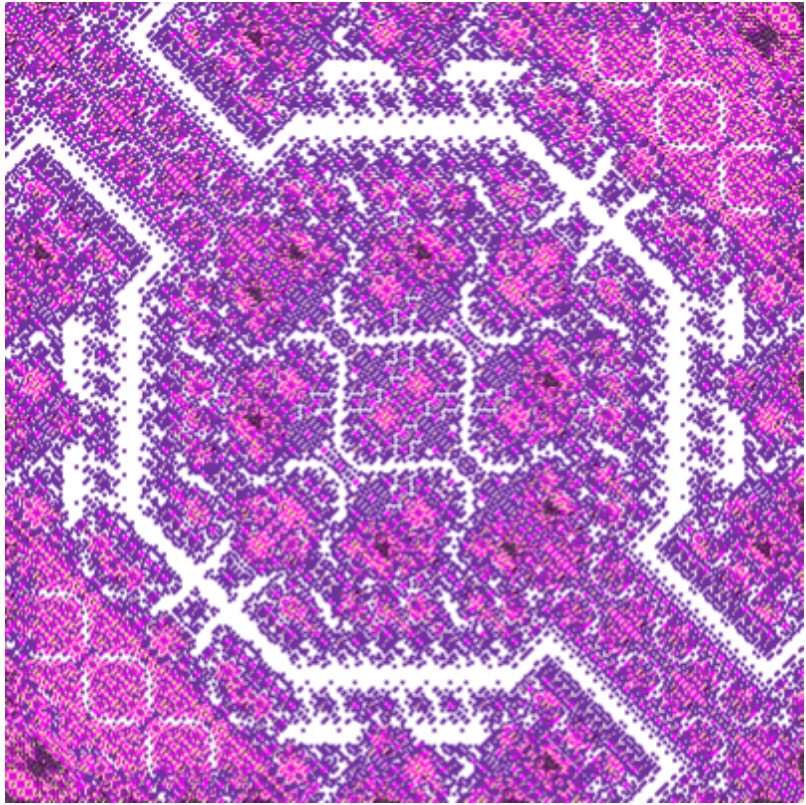}
\label{fig:401}
\end{figure}
\end{center}

\begin{paragraph}{\textbf{Acknowledgements}} I would like to thank Michael Baake, Peter Zeiner and Loh Jia Jun for helpful conversations. This research is funded by  D.~C.~O.\ was supported in part by a grant from the Fundamental Research Grant Scheme from the Malaysian Ministry of Education (with grant number FRGS/1/2022/TK07/XMU/01/1) and a Xiamen University Malaysia Research Fund (grant number: XMUMRF/2020-C5/IMAT/0011) 
\end{paragraph}
\end{section}
\begin{section}{Background}
\begin{subsection}{The Thue-Morse sequence}
The Thue-Morse sequence (sequence A010060 in the \emph{Online Encyclopedia of Integer Sequences \cite{ThueMorseOEIS}}) is a sequence of $a$'s and $b$'s that is defined in the following way.

Start with the length-$1$ sequence $\sigma_1=a$. Then to define $\sigma_j$ for $j\geq 2$, we replace every $a$ in the string  $\sigma_{j-1}$ with $ab$, and every $b$ in the string $\sigma_{j-1}$ with $ba$. For example, $\sigma_2=ab$ since we replaced the $a$ in $\sigma_1$ with $ab$. Then $\sigma_3=abba$ since we replaced the $a$ in $\sigma_2$ with $ab$, and we replaced the $b$ in $\sigma_2$ with $ba$. The first five $\sigma_j$ can then be defined similarly as follows:

\begin{align}
\sigma_1=&a\nonumber\\
\sigma_2=&ab\nonumber\\
\sigma_3=&abba\\
\sigma_4=&abbabaab\nonumber\\
\sigma_5=&abbabaabbaababba\nonumber\\
\ldots&\nonumber
\end{align}

We observe that the first entries of all the $\sigma_j$ are the same. This is easily proved by induction. This observation ensures that the following definition is well-defined:

\begin{definition}[Thue-Morse sequence]
The Thue-Morse sequence is an infinite binary sequence $\sigma$ such that for any $k\in \mathbb Z_+$, the $k$th entry of $\sigma$ is the same as the $k$th entry of $\sigma_j$, for all $\sigma_j$ that have length at least $k$.  
\end{definition} 
Thus 
\begin{equation}
\sigma=abbabaabbaababba\ldots
\end{equation}
See \cite{baakegrimm}, and \cite{AS99} for more extensive discussion on this Thue-Morse sequence.
\end{subsection}
\begin{subsection}{Thue-Morse Autocorrelation functions}
In this subsection, we introduce the autocorrelation function $\eta:\mathbb Z\to \mathbb Q$. This function arises from the mathematical theory of diffraction. To give a brief description of the physics. We are interested in how the aperiodically ordered lattice structure of a quasicrystal affects the diffraction of light that is shone through it. the Thue-Morse sequence is used as a 1-dimensional analogue of a quasicrystal lattice. The diffraction pattern of the quasicrystal is given by the Fourier transform of the autocorrelation measure of the Thue-Morse sequence, and this autocorrelation measure is a measure whose weights are determined by the autocorrelation function of the Thue-Morse sequence. In other words, the autocorrelation measure is $\sum_{m\in\mathbb Z} \eta(m)\delta_m$, where $\delta_m$ refers to a Dirac Delta at the location $m$. Details of this construction can be found in Section 9 of \cite{baakegrimm}.

In this paper, whenever we discuss the $n$th entry of a string of symbols, we start counting from $0$, i.e. the first entry is the $0$th entry. Now for $n\in\mathbb Z_{\geq 0}$ let us define
\begin{equation}\label{sigma}
 \sigma(n)=\begin{cases}
 +1, \text{ when the $n$th entry of the Thue-Morse sequence is $a$}\\
 -1, \text{ when the $n$th entry of the Thue-Morse sequence is $b$} 
 \end{cases}  
 \end{equation}
 Now we can define the Thue-Morse autocorrelation function.
 \begin{definition}\label{defn:eta}[Thue-Morse autocorrelation function]
 For $m\in \mathbb Z_{\geq 0}$, the Thue-Morse autocorrelation function is defined to be
 $$\eta(m)=\lim_{N\to\infty}\frac{1}{N}\sum_{i=0}^{N-1}\sigma(i)\sigma(i+m).$$
 This limit exists for any $m\in\mathbb Z_+$ (see the discussion around (2.2) of \cite{BaakeCoons}).
If $m$ is in $\mathbb Z_{<0}$, then we define $\eta(m)=\eta(-m)$. 
  
 \end{definition}
 It is not too hard to prove the following:
\begin{prop}[\cite{Kak72}]
 For all $m\in\mathbb Z_{\geq 0}$,
 $$\eta(2m)=\eta(m), \hspace{10pt}\eta(2m+1)=-\frac{1}{2}(\eta(m)+\eta(m+1)).$$
\end{prop}
This proposition gives us an alternate way of calculating $\eta(m)$. We may set $\eta(0)=1$ and $\eta(1)=-1/3$ and use the recursion relations to define the other $\eta(m)$.
\end{subsection}
\begin{subsection}{Higher order autocorrelation}
We can modify the definition of Definition \ref{defn:eta} so we are considering products of three or more Thue-Morse terms instead. For example,

\begin{definition}\label{defn:3eta}[Order 3 Thue-Morse autocorrelation function]
 For $(m,n)\in \mathbb Z_{\geq 0}^2$, the order 3 Thue-Morse autocorrelation function is defined to be
 $$\eta(m,n)=\lim_{N\to\infty}\frac{1}{N}\sum_{i=0}^{N-1}\sigma(i)\sigma(i+m)\sigma(i+n).$$
\end{definition}
However, this definition is not very interesting! Corollary 4.2 of \cite{BaakeCoons} says that $\eta(m,n)=0$ for all $m$ and $n$. So we instead proceed to order 4 correlations.
\begin{definition}\label{defn:4eta}[Order 4 Thue-Morse autocorrelation function]
 For $(m,n,k)\in \mathbb Z_{\geq 0}^3$, the order 4 Thue-Morse autocorrelation function is defined to be
 $$\eta(m,n,k)=\lim_{N\to\infty}\frac{1}{N}\sum_{i=0}^{N-1}\sigma(i)\sigma(i+m)\sigma(i+n)\sigma(i+k).$$
\end{definition}
 Let us first verify that Definition \ref{defn:4eta} is not as trivial as Definition \ref{defn:3eta}. It is easy to check, for instance that 
\begin{equation} 
 \eta(m,m,k)=\eta(m,k,m)=\eta(k,m,m)=\eta(k).
\end{equation}

 We can also develop a recursion algorithm to calculate $\eta(m,n,k)$.
 
 \begin{prop}
$$\eta(m,n,k)=\frac{(-1)^{m+n+k}}{2}\left(\eta\left(\lfloor m/2\rfloor,\lfloor n/2\rfloor,\lfloor k/2\rfloor \right)+\eta\left(\lceil m/2\rceil,\lceil n/2\rceil,\lceil k/2\rceil \right)\right)$$
\end{prop}
\begin{remark}
A generalized version of the above proposition can be found in \cite{BaakeCoons}, where they discuss order $n$ autocorrelations. 
\end{remark}
\begin{proof}
We will use (4.14) of \cite{baakegrimm} which states
\begin{equation}
\sigma(2j)=\sigma(j)\text{ and } \sigma(2j+1)=-\sigma(j).
\end{equation}
We can rewrite this as
\begin{equation}
\sigma(2i+j)=(-1)^j\sigma(i+\lfloor j/2\rfloor )
\end{equation}
for all $j\in\mathbb Z_{\geq 0}$. By the Birkhoff Ergodic Theorem and the fact that all the infinite sums are absolute continuous (and therefore rearrangement is allowed) we can calculate
 \begin{align}
 &\eta(m,n,k)\\
 =&\lim_{N\to\infty}\frac{1}{N}\sum_{i=0}^{N-1}\sigma(i)\sigma(i+m)\sigma(i+n)\sigma(i+k)\\
 =& \lim_{N\to\infty}\frac{1}{N}\sum_{i=0, i \text{ even}}^{N-1}\sigma(i)\sigma(i+m)\sigma(i+n)\sigma(i+k)\\
 &+\lim_{N\to\infty}\frac{1}{N}\sum_{i=0, i \text{ odd}}^{N-1}\sigma(i)\sigma(i+m)\sigma(i+n)\sigma(i+k)\\
=& \lim_{N\to\infty}\frac{(-1)^{m+n+k}}{N}\sum_{i=0, i \text{ even}}^{N-1}\sigma(i/2)\sigma(i/2+\lfloor m/2\rfloor )\sigma(i/2+\lfloor n/2 \rfloor )\sigma(i/2+\lfloor  k/2\rfloor )\\
 &+\lim_{N\to\infty}\frac{(-1)^{m+n+k}}{N}\\
 &\times\sum_{i=0, i \text{ odd}}^{N-1}\sigma\left(\frac{i-1}{2}\right)\sigma\left (\frac{i-1}{2}+\left\lfloor \frac{m+1}{2}\right\rfloor\right)\sigma\left (\frac{i-1}{2}+\left\lfloor \frac{n+1}{2}\right\rfloor\right)\sigma\left (\frac{i-1}{2}+\left\lfloor \frac{k+1}{2}\right\rfloor\right)\\
 =& \lim_{N\to\infty}\frac{(-1)^{m+n+k}}{N}\sum_{j=0}^{\lfloor (N-1)/2\rfloor }\sigma(j)\sigma(j+\lfloor m/2\rfloor )\sigma(j+\lfloor n/2 \rfloor )\sigma(j+\lfloor  k/2\rfloor )\\
 &+\lim_{N\to\infty}\frac{(-1)^{m+n+k}}{N}\sum_{j=0}^{\lfloor (N-2)/2\rfloor}\sigma\left(j\right)\sigma\left (j+\left\lceil m/2\right\rceil\right)\sigma\left (j+\left\lceil n/2\right\rceil\right)\sigma\left (j+\left\lceil k/2\right\rceil\right)\\
=&\frac{(-1)^{m+n+k}}{2}\left(\eta(\lfloor m/2\rfloor,\lfloor n/2\rfloor,\lfloor k/2\rfloor)+\eta(\lceil m/2\rceil,\lceil n/2\rceil,\lceil k/2\rceil)\right) 
  \end{align}
\end{proof}

\begin{remark}
An example where these higher order correlation functions appear in the mathematical physics literature is in \cite{Luck1989}. One key object in that paper is the \emph{complex Lyapunov exponent} $\Omega(E)$. Luck considers a discrete Schrodinger operator acting on $\ell^2(\mathbb Z_{\geq 0})$,

\begin{equation}\label{SchrodOp}
(H\psi)_n:=-\psi_{j+1}-\psi_{j-1}+V_j\psi_j, j\geq 0
\end{equation}
treating $\psi_{-1}$ as $0$. Here $V_j= V\sigma(j)$, where $V$ is a positive number and $\sigma(j)$ is defined in \eqref{sigma}.
\end{remark}
When we choose $\psi$ to be a formal eigenvector of $H$ corresponding to an eigenvalue $E\in \mathbb R$, we can define
\begin{equation}
\Omega(E)=\lim_{N\to\infty} \frac{1}{N}\sum_{j=0}^{N-1}\frac{\psi_{j+1}}{\psi_j}.
\end{equation}

Using the Schr\"odinger equation $H\psi=E\psi$ and \eqref{SchrodOp}, we expand $\Omega(E)$ in the variable $V$:

$\Omega(E)= \Omega^{(0)}+\Omega^{(1)} +\omega^{(2)}+\ldots+\omega^{(n)}+\ldots $ 
where for every $n$ $\Omega^{(n)}$ is a multiple of $V^n$ but not $V^{n+1}$. Luck calculates $\Omega^{(0)}$, $\Omega^{(1)}$, $\Omega^{(2)}$ and $\Omega^{(3)}$. The calculations (2.23) and (2.26) in \cite{Luck1989} demonstrate respectively that $\Omega^{(2)}$ can be expressed in terms of the form $\eta(n)$ and $\Omega^{(3)}$ can be expressed in terms of the form $\eta(m,n)$. It can be analogously calculate that $\Omega^{(4)}$ can be expressed in terms of the form $\eta(m,n,k)$ and so on. 

This construction is useful, because Luck uses perturbative behaviour of $\Omega(E)$ to understand the appearance of spectral gaps at $E$ as the variable $V$ varies. 

In fact, the motivation for this project emerged as I was attempting to understand the behaviour of $\Omega(E)$ in finer detail than presented in \cite{Luck1989}. I was using Microsoft Excel to record values of $\eta(x,y,z)$ in order to find patterns. I would fix a $z$, and let $x$ and $y$ each vary from $0$ to $z$. The columns of the excel sheet would represent $x$, and the rows of the excel sheet would represent $y$, with $z$ fixed. I would then fill in the spreadsheet cell with the corresponding value of $\eta(x,y,z)$. In order to make it easier to process the data, I would colour-code the values in the excel sheet, for example making every cell with a $0$ in it light blue, every cell with a $-1/4$ in it red, and so on. While doing this I realised that the Excel spreadsheet was producing rather striking images, and this prompted my curiosity as to what pictures could be created if I used values of $z$ that were in the hundreds and thousands. This motivated the art in the following section.

\end{subsection}
\end{section}
\begin{figure}[h]\label{fig:excel}
\caption{Colour-coded excel screenshot corresponding to calculating $\eta(x,y,z)$ with $z=20$}
\includegraphics[scale=.4]{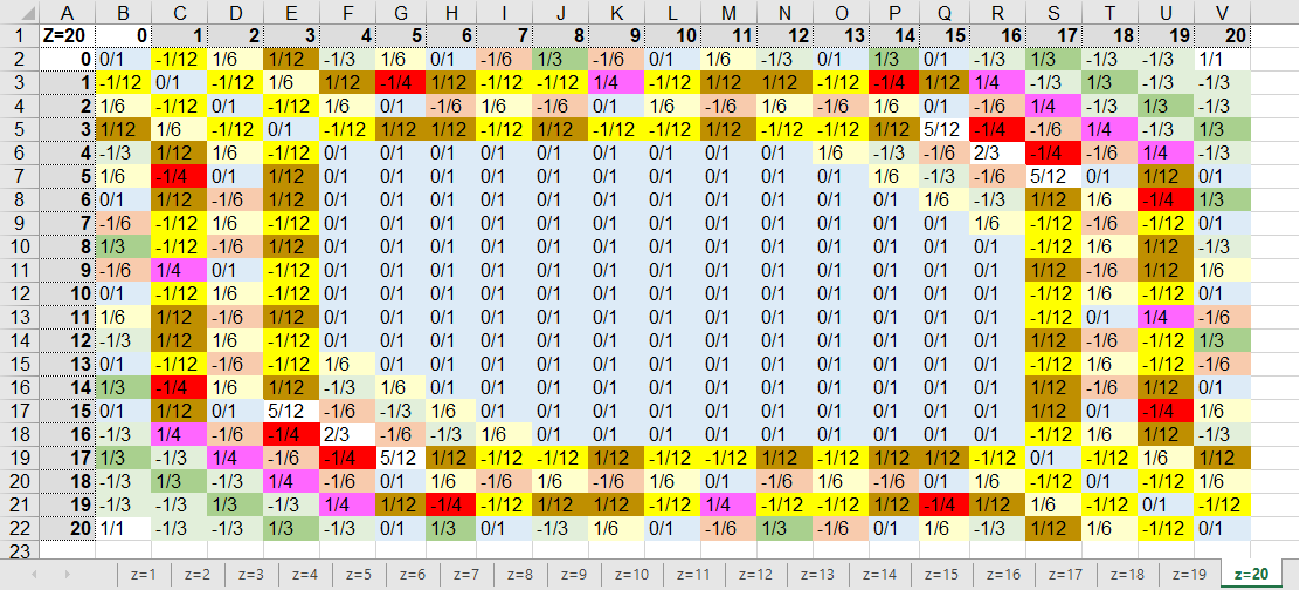} 
\end{figure}
\newpage
\begin{section}{Art generated from 4th Order Thue-Morse correlation functions}
The process for creating our Thue-Morse autocorrelation art is simply to extend the process in Figure \ref{fig:excel} to larger values of $z$. That is, fix $z$ to be a large integer. Then generate a $(z+1)\times (z+1) $ matrix where the entry in the $x$th row and $y$th column is $\eta(x,y,z)$ (here we start counting from the $x=0$th row and the $y=0$th column). Then create a color assignment function $f$ whose domain is $\mathbb R$ and whose codomain is a set of colors. We can then generate an image $z+1$ pixels wide and $z+1$ pixels high, where the pixel in $(x,y)$ has colour given by $f(\eta(x,y,z))$. Even for very simple functions $f$, this process can generate some striking images! We will demonstrate a few examples.

The following images are joint work with one of my undergraduate students, Loh Jia Jun. They were created using this colour assignment function. The output of the function is described using the format XXYYZZ in terms of RGB values, where XX represents the strength of the red component, YY represents the strength of the green component, and ZZ represents the strength of the blue component. The numbers XX, YY and ZZ are written in two-digit hexadecimal notation. Thus for example, FFFFFF represents black, 000000 represents white, and 00FF00 represents  {\color[HTML]{00FF00}green}. 

\begin{table}
\caption{The color assignment function}

\begin{center}
\begin{tabular}{|c|c|}
\hline
$x$&$f(x)$\\
\hline
$(-\infty,-0.1)$&{\color[HTML]{E8BA00}E8BA00}\\
\hline
$[-0.1,-0.05)$&{\color[HTML]{6C89EE}6C89EE}\\
\hline
$(-0.05,-0.025)$&{\color[HTML]{FFC750}FFC750}\\
\hline
$[-0.025,-0.01)$&{\color[HTML]{0000FF}0000FF}\\
\hline
$[-0.01,-0.008)$&{\color[HTML]{FFFF00}FFFF00}\\
\hline
$[-0.008,0)$&{\color[HTML]{56BEE9}56BEE9}\\
\hline
$(0,0.008]$&{\color[HTML]{83FE93}83FE93}\\
\hline
$(0.008,0.01]$&{\color[HTML]{FF0000}FF0000}\\
\hline
$(0.01,0.025]$&{\color[HTML]{00FF00}00FF00}\\
\hline
$(0.025,0.05]$&{\color[HTML]{00FFFF}00FFFF}\\
\hline
$(0.05,0.1]$&{\color[HTML]{84FF00}84FF00}\\
\hline
$(0.1,\infty)$&{\color[HTML]{FF765D}84FF00}\\
\hline
$\{-0.05,0\}$& white\\
\hline
\end{tabular} 
\end{center} 
\label{table:color}
\end{table}
The following images were created with the colour assignment function described above, with differing values for $z$.

\begin{figure}[h]
\caption{Image with $z=1023$}
\includegraphics[scale=0.3]{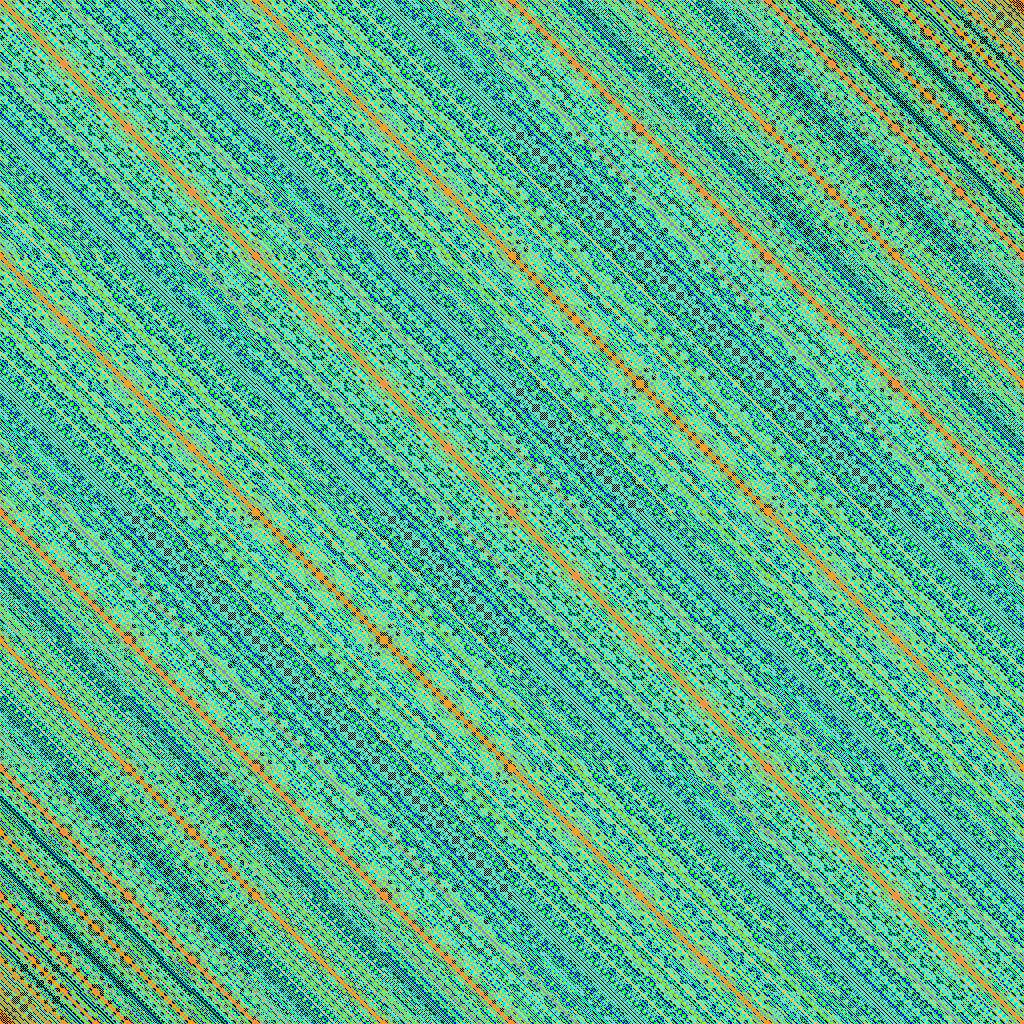}
\end{figure}
\newpage .

\begin{figure}[h]
\caption{Image with $z=1200$}
\includegraphics[scale=0.3]{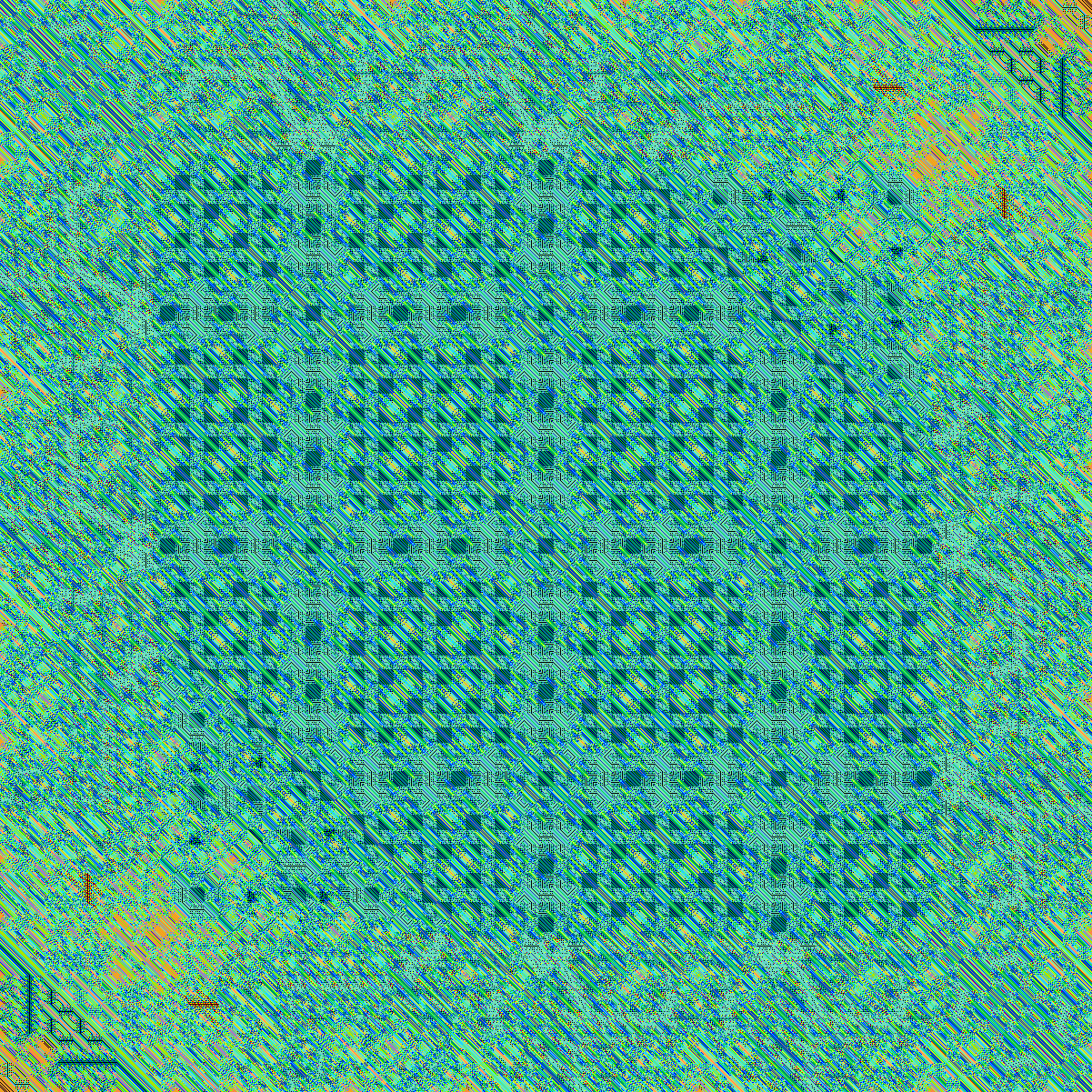}
\label{fig:1200}
\end{figure}
\newpage
.

\begin{figure}[H]
\caption{Image with $z=1475$}
\includegraphics[scale=0.3]{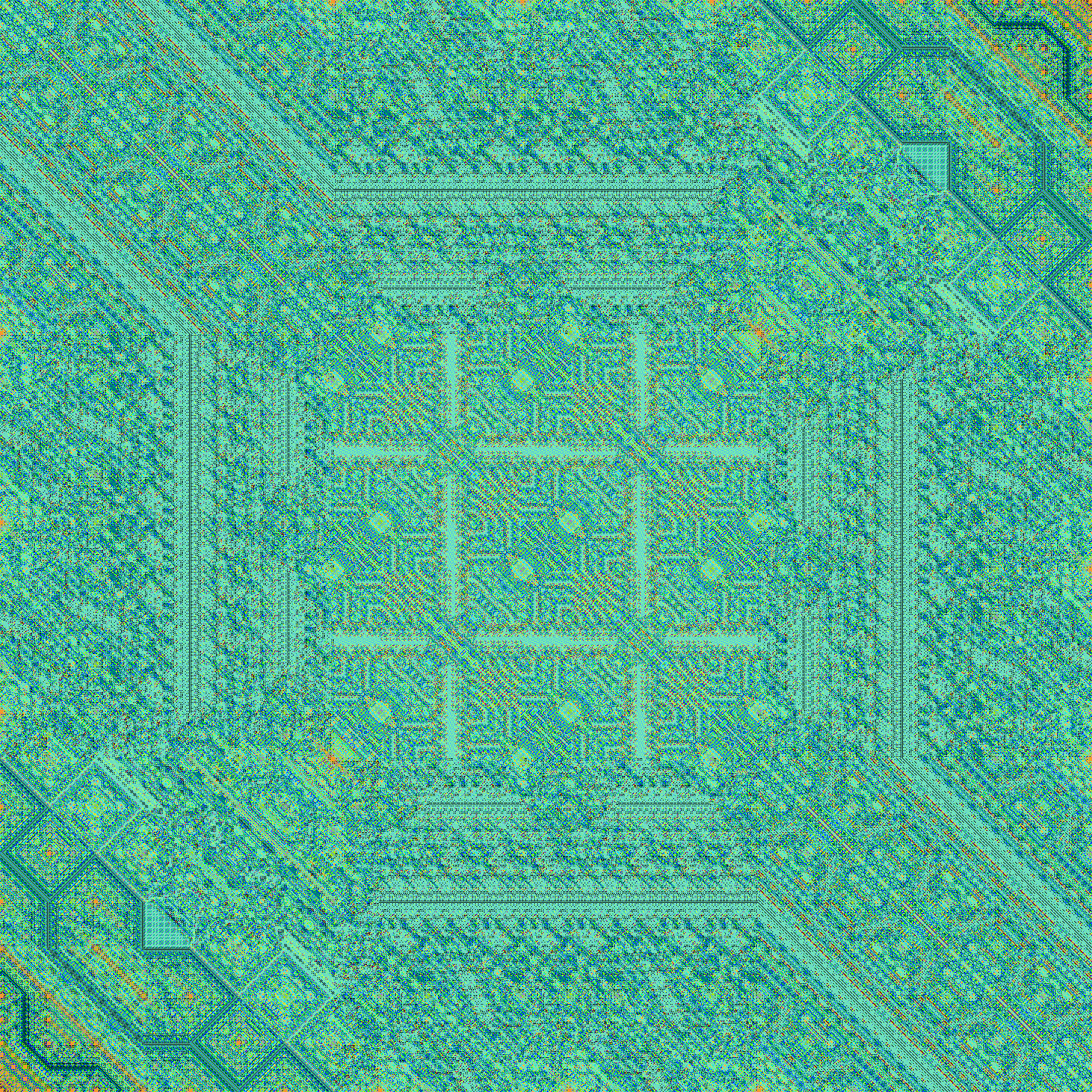}
\end{figure}

\end{section}
\newpage
\appendix
\begin{section}{Code}
The code to generate these images is written in Java 8. There are two separate programs. The first generates a .csv file, which is an array in which for a fixed $z$ the $m$th row and $n$th column is filled with the number $\eta(m,n,z)$ defined in Definition \ref{defn:4eta}. The latest version of the file is found in the github repository here: \url{https://doi.org/10.5281/zenodo.7060457}. To change the $z$-value, simply adjust the integer in line 19 of the program below.

\begin{lstlisting}[language=java, caption=Program to Generate .csv array]
import java.util.Scanner;
import java.io.FileNotFoundException;
import java.io.FileOutputStream;
import java.io.PrintStream;
import java.math.BigInteger;

	
public class TMcor {

	public static void main(String[] args) throws FileNotFoundException {
		final PrintStream oldStdout = System.out;
		System.setOut(new PrintStream(new FileOutputStream("ArtOutput.csv")));
		//Above gives the filename of the output file
		
		int[] input = new int[4];
		input[0]=1; //don't change this	
		input[1]=0;	//starting column, default is 0
		input[2]=0;	//starting row, default is 0
		input[3]=120;	//value of z
		for(int j=input[1];j<1+input[3];j++)
		{		String outstring="";
		
			for(int k=input[2];k<1+input[3];k++)
		{	
			int[] loopinput= new int[4];
			loopinput[0]=input[0];
			loopinput[1]=j;
			loopinput[2]=k;
			loopinput[3]=input[3];
			
	
			
			int[][][] output=tree(loopinput);
			int lastlevel=1+(int) Math.ceil(Math.log ((double) input[3])/Math.log(2));
			int oneplus=0;
			int oneminus=0;
			int zeroplus=0;
			int zerominus=0;
			for(int i=0;i<Math.pow(2,lastlevel-1);i++) {
				int sum =output[1][i][lastlevel-1]+output[2][i][lastlevel-1]+output[3][i][lastlevel-1];
			if(sum%2==0 && output[0][i][lastlevel-1]==1) zeroplus=zeroplus+1;
			if(sum%2==0 && output[0][i][lastlevel-1]==-1) zerominus=zerominus+1;
			if(sum%2==1 && output[0][i][lastlevel-1]==1) oneplus=oneplus+1;
			if(sum%2==1 && output[0][i][lastlevel-1]==-1) oneminus=oneminus+1;
			}
			int zerocalc=zeroplus-zerominus;
			int onecalc=oneplus-oneminus;
			
			double answer=(zerocalc*1.0-onecalc*(1.0/3))/Math.pow(2,lastlevel-1);
			outstring=outstring+""+ answer+",";
		
		} 	System.out.println(outstring);
		}
		System.setOut(oldStdout);
		System.out.println("The program ran successfully");
		}
		
		
	public static int[][] recursion (int[] start) {
		
		int[][] placeholder= new int[4][2];
placeholder[1][0]=(int) Math.floor(((float) start[1])/2);		
placeholder[2][0]=(int) Math.floor(((float) start[2])/2);
placeholder[3][0]=(int) Math.floor(((float) start[3])/2);	
placeholder[1][1]=(int) Math.ceil(((float) start[1])/2);		
placeholder[2][1]=(int) Math.ceil(((float) start[2])/2);
placeholder[3][1]=(int) Math.ceil(((float) start[3])/2);	
		
placeholder[0][0]=start[0];
if(placeholder[1][0]!=placeholder[1][1]) placeholder[0][0]=placeholder[0][0]*(-1);
if(placeholder[2][0]!=placeholder[2][1]) placeholder[0][0]=placeholder[0][0]*(-1);
if(placeholder[3][0]!=placeholder[3][1]) placeholder[0][0]=placeholder[0][0]*(-1);

placeholder[0][1]=placeholder[0][0];
		return placeholder;
	}
	
	//The output of tree is [sign,x,y,z][horizontal level][vertical level]
	public static int[][][] tree (int[] root){
		int levels=1+(int) Math.ceil(Math.log ((double) root[3])/Math.log(2));
		int[][][] x=new int[4][(int) Math.pow(2,levels-1)][levels];
		for(int i=0;i<4;i++) x[i][0][0]=root[i];
		for(int ell=1; ell<levels;ell++) {
		int horz=(int)Math.pow(2,ell);
		for(int j=0;j<horz/2;j++) {
			int[] temp1= new int[4];
			int[][] temp2= new int[4][2];
			for(int i=0;i<4;i++)temp1[i]=x[i][j][ell-1];
			temp2=recursion(temp1);
			for(int i=0;i<4;i++) {
		x[i][2*j][ell]=temp2[i][0];
		x[i][2*j+1][ell]=temp2[i][1];
			}
			}
		}
		
		
		return x;
	}
	}\label{listing:CSV}
\end{lstlisting}

The second program assigns colors to each entry in the array according to the function in Table \ref{table:color}. To change the color assignments, adjust the code from lines  65 to 130.

\begin{lstlisting}[language=java, caption=Program to generate colors from the .csv array]
import java.awt.image.BufferedImage;
import java.util.*;

import javax.imageio.ImageIO;

import java.lang.*;
import java.io.*;
 
public class TMart
{
	public static void main (String[] args) throws java.lang.Exception
	{
		
		
		   String csvFile = "ArtOutput.csv";
	        BufferedReader br = null;
	        String line = "";
	        String cvsSplitBy = ",";


	        BufferedReader brcount = new BufferedReader(new FileReader(csvFile));
	        int count = 0;
	        while(brcount.readLine() != null)
	        {
	            count++;
	        }
			brcount.close();
	  
	        double[][] DoubleMatrix= new double[count][count];
	        try {
	        	int row= 0; 
	            br = new BufferedReader(new FileReader(csvFile));
	            while ((line = br.readLine()) != null) {

	                // use comma as separator
	                String[] country = line.split(cvsSplitBy);

	                for(int i=0;i<count;i++) DoubleMatrix[row][i]=Double.parseDouble(country[i]);
	                row=row+1;
	            }

	        } catch (FileNotFoundException e) {
	            e.printStackTrace();
	        } catch (IOException e) {
	            e.printStackTrace();
	        } finally {
	            if (br != null) {
	                try {
	                    br.close();
	                } catch (IOException e) {
	                    e.printStackTrace();
	                }
	            }
	        }
	

        int[][] r = new int[count][count];
        int[][] g = new int[count][count];
        int[][] b = new int[count][count];
        
        for(int j=0; j<count;j++) 
        {
        	for (int k=0;k<count;k++) {
        		
        		if(DoubleMatrix[j][k]>0.008 && DoubleMatrix[j][k]<=0.01) {
        			r[j][k]=0xFF;
        			g[j][k]=0x00;
        			b[j][k]=0x00;
        			}
        		else if(DoubleMatrix[j][k]<-0.008&& DoubleMatrix[j][k]>=-0.01) {
        			r[j][k]=0xFF;
        			g[j][k]=0xFF;
        			b[j][k]=0x00;
        			}
        		else if(DoubleMatrix[j][k]>0.01 && DoubleMatrix[j][k]<=0.025) {
        			r[j][k]=0x00;
        			g[j][k]=0xFF;
        			b[j][k]=0x00;
        		}
        		else if(DoubleMatrix[j][k]<-0.01 && DoubleMatrix[j][k]>=-0.025) {
        			r[j][k]=0x00;
        			g[j][k]=0x00;
        			b[j][k]=0xFF;
        		}
        		else if(DoubleMatrix[j][k]>0.025 && DoubleMatrix[j][k]<=0.05) {
        			r[j][k]=0x00;
        			g[j][k]=0xFF;
        			b[j][k]=0xFF;
        		}
        		else if(DoubleMatrix[j][k]<-0.025 && DoubleMatrix[j][k]>-0.05) {
        			r[j][k]=0xFF;
        			g[j][k]=0xC7;
        			b[j][k]=0x50;
        		}
        		else if(DoubleMatrix[j][k]>0 && DoubleMatrix[j][k]<=0.008) {
        			r[j][k]=0x83;
        			g[j][k]=0xFE;
        			b[j][k]=0x93;
        		}
        		else if(DoubleMatrix[j][k]<0 && DoubleMatrix[j][k]>=-0.008) {
        			r[j][k]=0x56;
        			g[j][k]=0xBE;
        			b[j][k]=0xE9;
        		}
        		else if(DoubleMatrix[j][k]<-0.05&& DoubleMatrix[j][k]>=-0.1) {
        			r[j][k]=0x6C;
        			g[j][k]=0x89;
        			b[j][k]=0xEE;
        			}
        		
        		else if(DoubleMatrix[j][k]>0.05 && DoubleMatrix[j][k]<=0.1) {
        			r[j][k]=0x84;
        			g[j][k]=0xFF;
        			b[j][k]=0x00;
        			}
        		else if(DoubleMatrix[j][k]<-0.1) {
        			r[j][k]=0xE8;
        			g[j][k]=0xBA;
        			b[j][k]=0x00;
        			}
        		else if(DoubleMatrix[j][k]>0.1) {
        			r[j][k]=0xFF;
        			g[j][k]=0x76;
        			b[j][k]=0x5D;
        			}
        		else {
        			r[j][k]=0x00;
        			g[j][k]=0x00;
        			b[j][k]=0x00;
        			}
        		
        	}
        	
        	
        }
        
        int width = count;
        int height = count;
 
        BufferedImage image = new BufferedImage(width, height, BufferedImage.TYPE_INT_RGB); 
 
        for (int y = 0; y < height; y++) {
            for (int x = 0; x < width; x++) {
                int rgb = r[y][x];
                rgb = (rgb << 8) + g[y][x]; 
                rgb = (rgb << 8) + b[y][x];
                image.setRGB(x, y, rgb);
            }
        }
 
  
        
        File outputFile = new File("output.bmp");
         ImageIO.write(image, "bmp", outputFile);
    }
}\label{listing:color}
\end{lstlisting}
Running these two programs in succession without changes will generate the image in Figure \ref{fig:1200}.

\subsection{Simple Example}
Let us go through a simple example. We modify line 19 of Listing 1 to set \texttt{input[3]}$=3$. When we run the program, it will generate the following .csv file:

\begin{figure}[h]\label{fig:excel3}
\caption{ArtOutput.csv file generated by Listing 1 with \texttt{input[3]}$=3$}
\includegraphics[scale=.4]{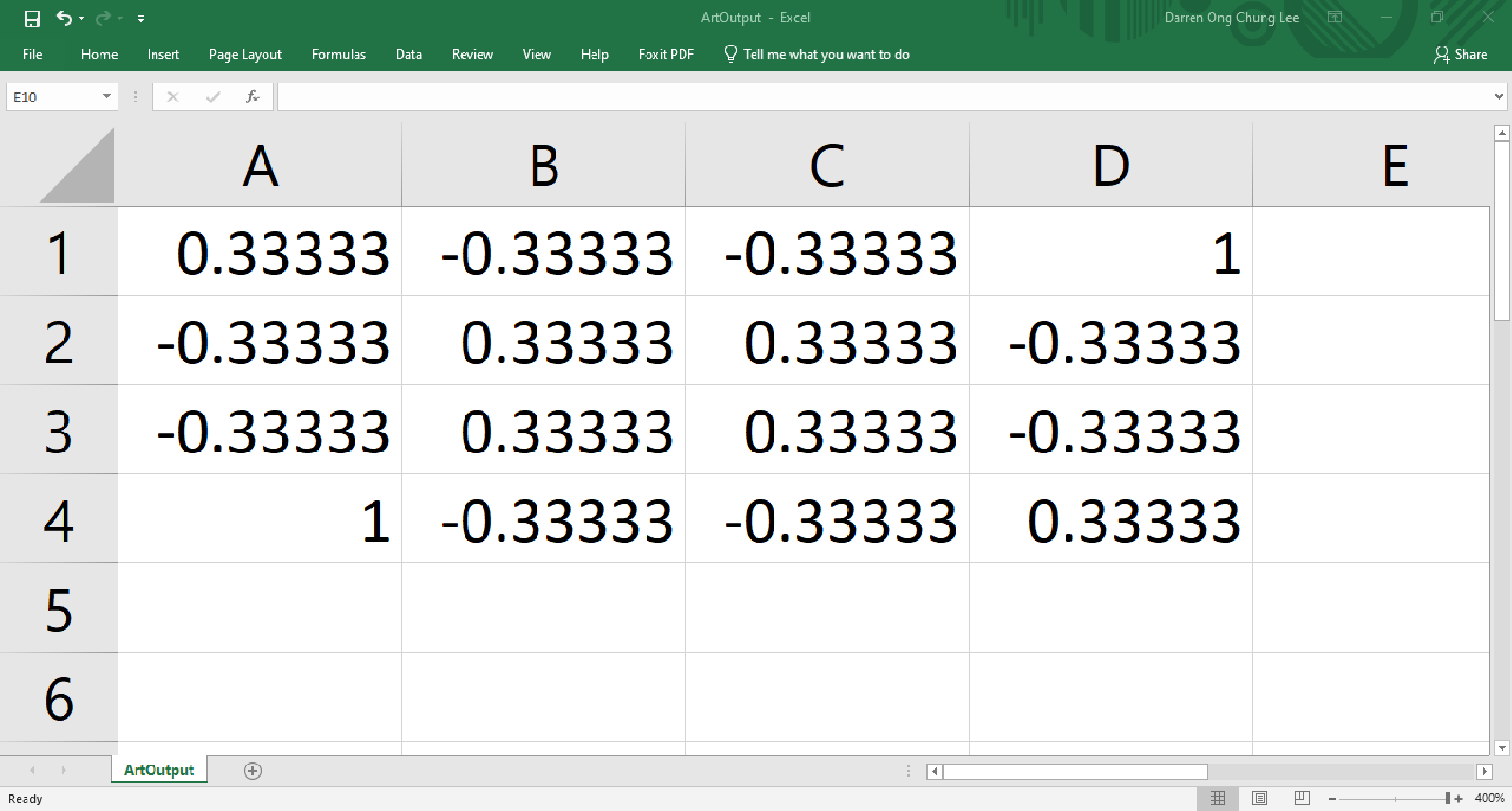} 
\end{figure}

Then, using this .csv file as the input for the program in Listing 2, we obtain the following image:

\begin{figure}[h]\label{fig:3pic}
\caption{.bmp image file generated by Listing 2 using the .csv file in Figure \ref{fig:excel3} as input.}
\includegraphics[scale=3]{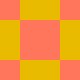} 
\end{figure}

\end{section}
\bibliographystyle{alpha}
\bibliography{darren}

\begin{thebibliography}{ATR14}

\bibitem[AA12]{alajlouni2012}
Rima~A Al~Ajlouni.
\newblock The global long-range order of quasi-periodic patterns in islamic
  architecture.
\newblock {\em Acta Crystallographica Section A: Foundations of
  Crystallography}, 68(2):235--243, 2012.

\bibitem[AS99]{AS99}
J.-P. Allouche and J.~Shallit.
\newblock The ubiquitous {P}rouhet-{T}hue-{M}orse sequence.
\newblock In C.~Ding, T.~Helleseth, and H.~Niederreiter, editors, {\em
  Sequences and Their Applications: Proceedings of {S}{E}{T}{A} ’98}, pages
  1--16. Springer Berlin, 1999.

\bibitem[ATR14]{ATR2014}
Youssef Aboufadil, Abdelmalek Thalal, and My~Ahmed El~Idrissi Raghni.
\newblock Moroccan ornamental quasiperiodic patterns constructed by the
  multigrid method.
\newblock {\em Journal of Applied Crystallography}, 47(2):630--641, 2014.

\bibitem[BC22]{BaakeCoons}
Michael Baake and Michael Coons.
\newblock Correlations of the {T}hue-{M}orse sequence.
\newblock {\em arXiv:2209.07102}, 2022.

\bibitem[BG13]{baakegrimm}
Michael Baake and Uwe Grimm.
\newblock {\em Aperiodic Order: Volume 1, A Mathematical Invitation}, volume
  149 of {\em Encyclopedia of Mathematics and its Applications}.
\newblock Cambridge University Press, 2013.

\bibitem[BG17]{baakegrimm2}
Michael Baake and Uwe Grimm.
\newblock {\em Aperiodic Order: Volume 2, Crystallography and Almost
  Periodicity}, volume 166 of {\em Encyclopedia of Mathematics and its
  Applications}.
\newblock Cambridge University Press, 2017.

\bibitem[Brl89]{Brlek89}
Sre{\'c}ko Brlek.
\newblock Enumeration of factors in the {T}hue-{M}orse word.
\newblock {\em Discrete Applied Mathematics}, 24(1-3):83--96, 1989.

\bibitem[Fou22]{ThueMorseOEIS}
OEIS Foundation.
\newblock {O}nline {E}ncyclopedia of {I}nteger {S}equences:
  {A}010060-{O}{E}{I}{S}.
\newblock \url{https://oeis.org/A010060}, 2022.

\bibitem[Kak72]{Kak72}
S.~Kakutani.
\newblock Strictly ergodic symbolic dynamical systems.
\newblock In L.~M.~Le Cam, J.~Neyman, and E.L. Scott, editors, {\em Proceedings
  of the Sixth Berkeley Symposium on Mathematical Statistics and Probability},
  pages 319--326. University of California Press, 1972.

\bibitem[LS07]{LuSteinhardt2007}
Peter~J Lu and Paul~J Steinhardt.
\newblock Decagonal and quasi-crystalline tilings in medieval {I}slamic
  architecture.
\newblock {\em Science}, 315(5815):1106--1110, 2007.

\bibitem[Luc89]{Luck1989}
JM~Luck.
\newblock Cantor spectra and scaling of gap widths in deterministic aperiodic
  systems.
\newblock {\em Physical Review B}, 39(9):5834, 1989.

\bibitem[L{\"u}c00]{luck2000}
Reinhard L{\"u}ck.
\newblock D{\"u}rer--{K}epler--{P}enrose, the development of pentagon tilings.
\newblock {\em Materials Science and Engineering: A}, 294:263--267, 2000.

\bibitem[Moo97]{Moody1997}
Robert~V Moody.
\newblock {\em The Mathematics of Long-Range Aperiodic Order}, volume 489 of
  {\em NATO Science Series C}.
\newblock Springer, 1997.

\bibitem[Ong20]{Ong2020}
Darren~C Ong.
\newblock Quasiperiodic music.
\newblock {\em Journal of Mathematics and the Arts}, 14(4):285--296, 2020.

\bibitem[Pru86]{Prusinkiewicz1986}
Przemyslaw Prusinkiewicz.
\newblock Score generation with {L}-systems.
\newblock In {\em Proceedings of the 1986 International Computer Music
  Conference}, pages 455--457, 1986.

\bibitem[Tre22]{Trevino2022}
Rodrigo Trevi{\~n}o.
\newblock Quasimusic: tilings and metre.
\newblock {\em Journal of Mathematics and the Arts}, 16(1-2):162--181, 2022.

\end{thebibliography}
\end{document}